\documentclass[sigconf]{acmart}

\usepackage{algorithm}
\usepackage{algorithmic}
\usepackage{multirow}
\usepackage{amsmath}
\usepackage{amsthm}
\usepackage{enumitem}
\setlist[itemize]{leftmargin=*}
\usepackage{booktabs}
\usepackage{graphicx}
\usepackage{subcaption}
\usepackage{balance}
\usepackage{microtype}
\usepackage{float}

\AtBeginDocument{
  }

\copyrightyear{2026}
\acmYear{2026}
\setcopyright{cc}
\setcctype{by}
\acmDOI{10.1145/3770855.3818363}

\acmConference[KDD '26]{Proceedings of the 32nd ACM SIGKDD Conference on Knowledge Discovery and Data Mining V.2}{August 9--13, 2026}{Jeju Island, Republic of Korea}
\acmBooktitle{Proceedings of the 32nd ACM SIGKDD Conference on Knowledge Discovery and Data Mining V.2 (KDD '26), August 9--13, 2026, Jeju Island, Republic of Korea}
\acmISBN{979-8-4007-2259-2/2026/08}

\begin{document}

\title{DeGRe: Dense-supervised Generative Reranking for Recommendation}

\author{Chaotian Song}
\authornote{Work done during an internship at Taobao Shangou, Alibaba Group.}
\affiliation{
  \institution{College of Software, Zhejiang~University}
  \city{Hangzhou}
  \country{China}
}
\email{songchaotian@zju.edu.cn}

\author{Jingyao Zhang}
\affiliation{
  \institution{Rajax Network Technology, \\ Taobao Shangou of Alibaba}
  \city{Hangzhou}
  \country{China}
}
\email{xingshu.zjy@alibaba-inc.com}

\author{Chenghao Chen}
\affiliation{
  \institution{Rajax Network Technology, \\ Taobao Shangou of Alibaba}
  \city{Hangzhou}
  \country{China}
}
\email{chenchenghao.cch@alibaba-inc.com}

\author{Zisen Sang}
\affiliation{
  \institution{Rajax Network Technology, \\ Taobao Shangou of Alibaba}
  \city{Beijing}
  \country{China}
}
\email{zisen.szs@alibaba-inc.com}

\author{Dehai Zhao}
\affiliation{
  \institution{College of Software, Zhejiang~University}
  \city{Hangzhou}
  \country{China}
}
\email{dehai.zhao@zju.edu.cn}

\author{Guodong Cao}
\affiliation{
  \institution{Rajax Network Technology, \\ Taobao Shangou of Alibaba}
  \city{Beijing}
  \country{China}
}
\email{guodong.cao@alibaba-inc.com}

\author{Boxi Wu}
\authornote{Corresponding author.}
\affiliation{
  \institution{College of Software, Zhejiang~University}
  \city{Hangzhou}
  \country{China}
}
\email{wuboxi@zju.edu.cn}

\author{Deng Cai}
\affiliation{
  \institution{State Key Lab of CAD\&CG, Zhejiang~University}
  \city{Hangzhou}
  \country{China}
}
\email{dengcai@cad.zju.edu.cn}

\author{Jia Jia}
\affiliation{
  \institution{Rajax Network Technology, \\ Taobao Shangou of Alibaba}
  \city{Shanghai}
  \country{China}
}
\email{jj229618@alibaba-inc.com}

\renewcommand{\shortauthors}{Chaotian Song et al.}

\begin{abstract}
In multi-stage recommender systems, reranking optimizes overall utility by capturing intra-list contextual dependencies, yet its central challenge lies in exploring optimal sequences within an exponentially large permutation space. Recent studies have shifted towards end-to-end generative frameworks, which typically leverage list--wise rewards or preference alignment to guide generator training. However, these methods still face two critical issues. First is the heuristic label bias. Existing methods often construct training targets based on simple rules, such as promoting clicked items to the top, while ignoring causal dependencies within the list context. Second is the credit assignment problem. Sparse list-level posterior rewards fail to directly guide intermediate steps in sequence generation, leading to ambiguous optimization directions.

To address these issues, we propose DeGRe (\textbf{De}nse-supervised \textbf{G}enerative \textbf{Re}ranking), a generative reranking framework that bridges the gap between offline exploration and online efficiency through dense supervision. The core of DeGRe lies in its offline–online decoupled design. During the offline phase, we introduce a Lookahead Evaluator based on cumulative regression, which leverages beam search to actively mine high-value lookahead sequences in the unexposed space. During training, we transform the step--wise value estimations from the evaluator into dense supervision signals and distill them into a lightweight Online Generator. This mechanism enables the generator to internalize lookahead planning capabilities, requiring only a single efficient greedy decoding pass during online inference to approximate the global optimum. Experiments demonstrate that DeGRe outperforms baseline models on public benchmarks and industrial datasets. We have successfully deployed DeGRe on Taobao Flash Shopping, significantly improving online recommendations.
\end{abstract}



\maketitle

\section{Introduction}

Large-scale e-commerce platforms such as Taobao provide personalized recommendation services to users from hundreds of millions of items on a daily basis. To balance efficiency and effectiveness, industrial systems commonly adopt a multi-stage recommendation architecture \cite{Covington2016YouTubeDNN}: recall, ranking, and reranking. The recall stage \cite{Huang2013DSSM, Zhu2018TDM, Li2019MIND} is responsible for filtering relevant items from a massive corpus, while the ranking stage \cite{Cheng2016WideDeep, Guo2017DeepFM, Wang2017DCN, Zhou2018DIN, Zhou2019DIEN, Jiang2025IAK} typically relies on point--wise score estimation. The reranking stage \cite{Zhuang2018miRNN, Bello2018Seq2Slate, Xi2022MIR}, positioned at the end of the pipeline, reorganizes the candidate set by modeling contextual associations among items to maximize overall utility \cite{Liu2022LibRerank}.

The central challenge in reranking lies in exploring optimal sequences within an exponentially large permutation space. Early single-stage greedy strategies \cite{Ai2018DLCM, Pei2019PRM} attempt to rescore candidates by encoding context, but the inherent limitation of evaluation-before-reranking \cite{Feng2021GRN, Xi2024URCC} often causes models to converge to suboptimal solutions. To overcome this bottleneck, the generator-evaluator paradigm \cite{Feng2021GRN, Feng2021PRS, Shi2023PIER, Lin2024DCDR, Ren2024NAR4Rec} has been widely adopted. In this paradigm, the generator produces multiple candidate lists, and the evaluator selects the best one. However, this two--stage architecture faces the critical problem of goal inconsistency \cite{Wang2025NLGR}: the evaluator aims to fit list-level value, whereas the generator must transform candidates into optimal sequences. The divergence in objectives and the structural separation hinder end-to-end optimization \cite{Lin2025GReF}. Moreover, deploying both the generator and the evaluator simultaneously increases system complexity and online latency. To address these limitations, recent research has shifted toward more efficient end-to-end generative reranking frameworks \cite{Wang2025NLGR, Zhang2025GoalRank, Lin2025GReF}. For example, GReF \cite{Lin2025GReF} achieves both inference efficiency and unified generation-evaluation through ordered multi-token prediction, while NLGR \cite{Wang2025NLGR} and GoalRank \cite{Zhang2025GoalRank} leverage reward signals to guide generator training.

\begin{figure}[t]
  \centering
  \includegraphics[width=\columnwidth]{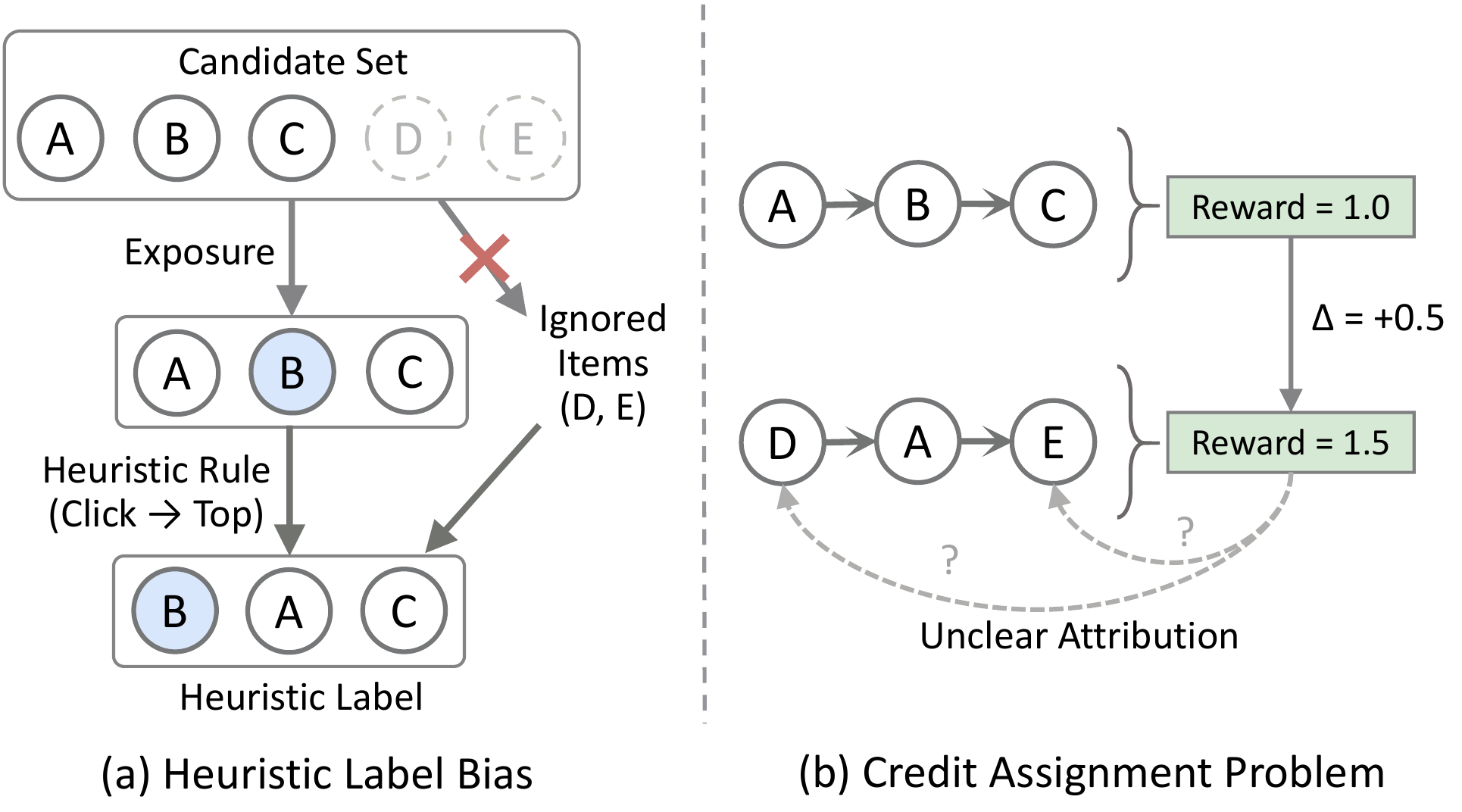}
  \caption{Illustration of two core challenges in existing generative reranking methods.}
  \label{fig:fig1}
\end{figure}

However, existing generative reranking methods still face two critical issues. The first is heuristic label bias. As illustrated in Figure~\ref{fig:fig1}(a), existing methods \cite{Lin2025GReF} often construct training labels based on heuristic rules, such as promoting clicked items to the top. This practice ignores the causal dependencies within the list context and implicitly assumes that clicked items yield higher value than non-clicked items regardless of position. Consequently, the model tends to fit a biased data distribution rather than learn the true globally optimal ranking. This not only introduces bias but also limits the ability of the model to mine potentially high-value sequences in the unexposed space. The second is the credit assignment problem. Although some works introduce reward models \cite{Wang2025NLGR, Zhang2025GoalRank}, they typically rely on list--wise posterior rewards, such as overall CTR or GMV. As shown in Figure~\ref{fig:fig1}(b), such coarse-grained scalar signals are difficult to attribute effectively to each local decision during sequence generation. Without fine-grained step--wise guidance, the generator cannot discern the specific contribution of each action, leading to ambiguous optimization directions and ultimately limiting the performance ceiling.

To address these issues, we propose DeGRe (\textbf{De}nse-supervised \textbf{G}enerative \textbf{Re}ranking). The core strategy is offline--online decoupling: offloading the computationally intensive exploration of the permutation space to the offline stage, thereby achieving both low online inference latency and high-quality generation. Specifically, during the offline phase, we introduce a Lookahead Evaluator based on cumulative regression that estimates the expected cumulative value after appending each candidate item. Using this evaluator, we employ beam search to explore the unexposed space and mine high-value lookahead sequences. During training, we transform the step--wise value estimations from the evaluator into dense supervision signals and distill them into the Online Generator. This mechanism enables the generator to internalize lookahead planning capabilities, allowing it to approximate the global optimum through a single efficient greedy decoding pass during online inference.

In summary, the main contributions of this paper are as follows:
\begin{itemize}
\item We propose DeGRe, a framework that adopts an offline--online decoupling strategy. This strategy enables the use of abundant offline computational resources to actively mine high-value sequences in the unexposed space, effectively mitigating bias introduced by heuristic labels.
\item We design a dense supervision mechanism based on the Lookahead Evaluator. Unlike methods that rely on sparse posterior rewards, our approach provides step--wise value estimations for dense guidance, effectively alleviating the credit assignment problem during generation.
\item Experiments on public and industrial datasets demonstrate that DeGRe outperforms existing state-of-the-art methods while maintaining efficient inference. Online A/B tests confirm significant business value (GMV improvement of +3.75\%). DeGRe is currently deployed in Taobao Flash Shopping.
\end{itemize}

\section{Related Work}

\subsection{Reranking in Recommendation}

Unlike recall and ranking, reranking focuses on modeling item correlations within the exposed list. Typical reranking methods fall into two categories. The first category consists of one--stage methods \cite{Ai2018DLCM, Pei2019PRM, Pang2020SetRank, Xi2022MIR, Chen2022EXTR}, which treat reranking as a context-aware scoring task. For example, DLCM \cite{Ai2018DLCM} and PRM \cite{Pei2019PRM} employ GRU \cite{Cho2014GRU, Chung2014GRU} and Transformer \cite{Vaswani2017Transformer}, respectively, to encode contextual information of the initial list and output revised scores for greedy sorting. However, such methods suffer from the inherent limitation of evaluation-before-reranking \cite{Xi2024URCC, Feng2021GRN}, which often leads to suboptimal results.

The other category includes two--stage methods \cite{Feng2021GRN, Feng2021PRS, Gong2022EdgeRerank, Shi2023PIER, Lin2024DCDR, Ren2024NAR4Rec}, which adopt a generator-evaluator framework. In this framework, the generator produces multiple candidate sequences, while the evaluator selects the optimal sequence based on the estimated list value. Although this paradigm outperforms one--stage methods in effectiveness, it faces two major challenges: the separation between the generator and the evaluator leads to the problem of goal inconsistency \cite{Wang2025NLGR}; and the need to sample and re-score a large number of candidate sequences incurs high computational complexity, limiting its applicability in real-time industrial systems.

To overcome these limitations, recent research has shifted toward more efficient end-to-end generative reranking frameworks. GReF \cite{Lin2025GReF} unifies generation and evaluation through ordered multi-token prediction (OMTP), while NLGR \cite{Wang2025NLGR} and GoalRank \cite{Zhang2025GoalRank} leverage reward signals to guide generator training. However, these methods still face the challenges mentioned earlier: MLE-based methods \cite{Lin2025GReF} essentially fit the biased distribution introduced by heuristic labels, making it difficult to explore high-value sequences in the unexposed space. Reward-based methods \cite{Wang2025NLGR, Zhang2025GoalRank} are constrained by sparse posterior feedback and lack fine-grained guidance for intermediate generation steps.

\subsection{List Value Estimation}

Accurate estimation of list value is crucial for reranking. From early point--wise models \cite{Cheng2016WideDeep, Lian2018xDeepFM, Zhou2018DIN} to list--wise models \cite{Pei2019PRM, Shi2023PIER, Wang2025NLGR}, the modeling of contextual dependencies in reranking has become increasingly precise. NLGR \cite{Wang2025NLGR} and GoalRank \cite{Zhang2025GoalRank} further employ these models as reward functions. However, existing list--wise evaluators only score complete lists, making it difficult to directly guide step--wise decisions during autoregressive generation. Cumulative regression \cite{Frank2001Ordinal}, by modeling the cumulative distribution of discrete metrics, is naturally suited for estimating the progressive value of subsequences. Building on this, we employ cumulative regression to construct a Lookahead Evaluator, providing the generator with fine-grained value signals at each step.

\begin{figure*}[t]
  \centering
  \includegraphics[width=\linewidth,height=0.34\textheight,keepaspectratio]{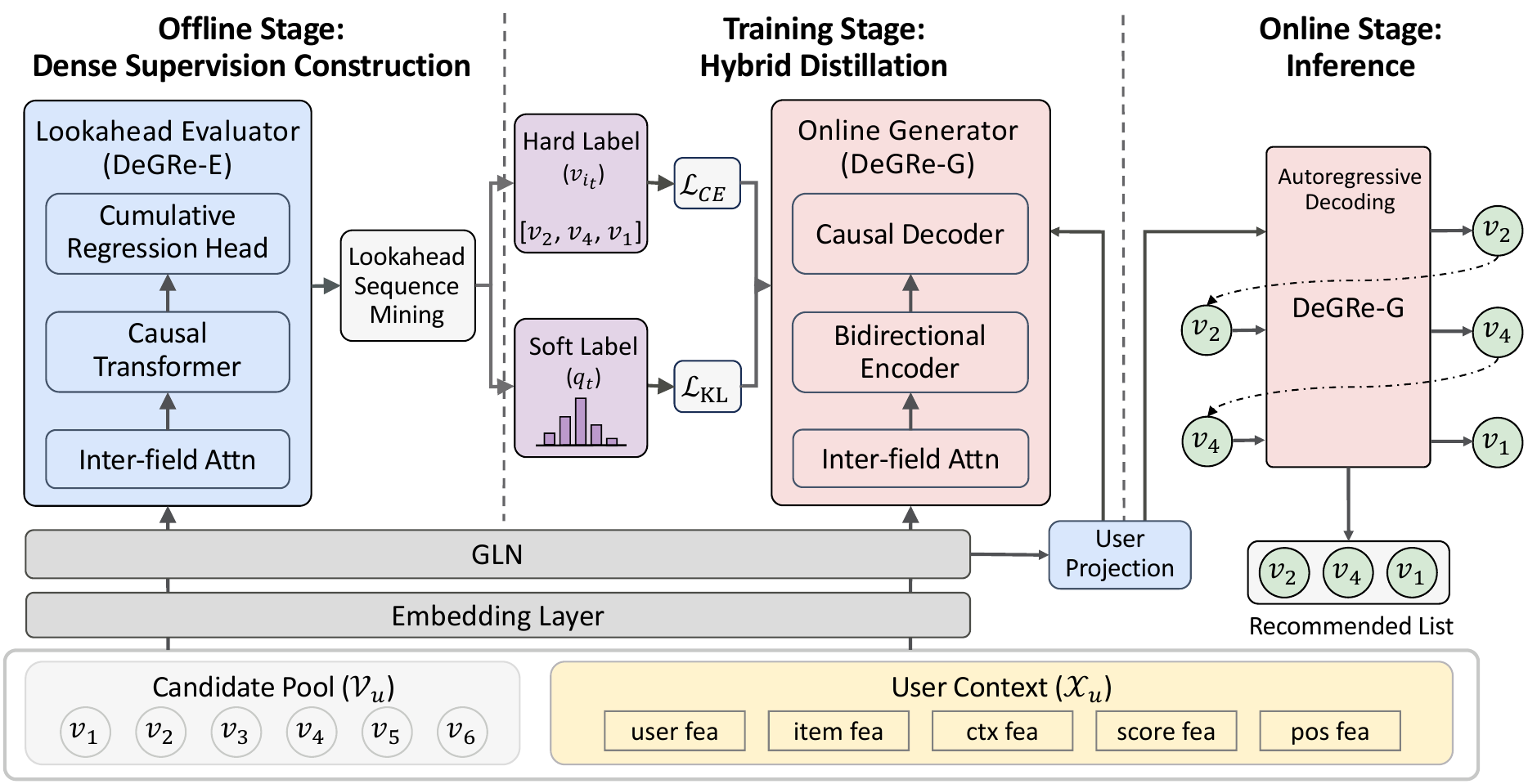}
  \caption{Overall framework of DeGRe. In the offline phase, the Lookahead Evaluator constructs dense supervision signals through lookahead sequence mining. In the training phase, the Online Generator internalizes lookahead planning capabilities via hybrid distillation. In the online phase, the model achieves efficient inference with a single greedy decoding pass.}
  \label{fig:fig2}
\end{figure*}

\section{Preliminaries}

\subsection{Problem Formulation}

In the multi-stage recommendation architecture, reranking is positioned after the ranking stage. For each user request $u \in \mathcal{U}$, the system provides a candidate set $\mathcal{V}_u = \{v_1, v_2, \dots, v_N\}$ containing $N$ items along with the corresponding user context features $\mathcal{X}_u$. The core task of reranking is to select $L$ items from $\mathcal{V}_u$ (where $L < N$) and determine their display order, thereby generating the final recommendation list $l = [v_{i_1}, v_{i_2}, \dots, v_{i_L}]$, where each $v_{i_k} \in \mathcal{V}_u$ and all items are distinct. We denote the space of all possible permutations of length $L$ as $\Pi(\mathcal{V}_u)$. Our objective is to find an optimal list $l^* \in \Pi(\mathcal{V}_u)$ that maximizes the overall utility $V(l | \mathcal{X}_u)$ (e.g., total clicks):
\begin{equation}
l^* = \arg\max_{l \in \Pi(\mathcal{V}_u)} V(l | \mathcal{X}_u)
\end{equation}

\textbf{Generative Reranking.} To capture combinatorial effects within the list, generative reranking formulates the task as a sequential decision process. Based on the probability chain rule, the probability of generating list $l$ is decomposed into a product of conditional probabilities at each step:
\begin{equation}
P_\theta(l | \mathcal{V}_u, \mathcal{X}_u) = \prod_{t=1}^L P_\theta(v_{i_t} | l_{<t}, \mathcal{V}_u \setminus l_{<t}, \mathcal{X}_u)
\end{equation}
where $l_{<t} = [v_{i_1}, \dots, v_{i_{t-1}}]$ denotes the subsequence generated before step $t$.

\subsection{Cumulative Regression}

To provide dense intermediate feedback for the generation process, we need to accurately estimate the actual cumulative value of subsequences at arbitrary lengths. We adopt cumulative regression~\cite{McCullagh1980Ordinal} to model discrete value distributions. Given a subsequence $l_{1:t}$, instead of directly regressing its scalar value as in traditional approaches, we model the distribution of the current cumulative value $V$ (e.g., the total number of clicks obtained so far), that is, predicting the conditional probability that $V$ reaches or exceeds a threshold $k$:
\begin{equation}
P(V \ge k | l_{1:t}) = \sigma(f_k(l_{1:t})), \quad \forall k \in \{1, \dots, t\}
\end{equation}
where $f_k \in \mathbb{R}$ is the logit output corresponding to threshold $k$. Intuitively, at any time step $t$, the model outputs a vector of dimension $t$, where the $k$-th element represents the confidence that the current value is at least $k$. This design essentially decomposes the regression task into a set of ordinal binary classification problems~\cite{Frank2001Ordinal}. The expected cumulative value of a subsequence can be directly derived by summing these probabilities:
\begin{equation}
\mathbb{E}[V | l_{1:t}] = \sum_{k=1}^t P(V \ge k | l_{1:t})
\end{equation}

\section{Methodology}

Figure~\ref{fig:fig2} presents the overall architecture of DeGRe. Formally, given the user context $\mathcal{X}_u$ and a candidate set $\mathcal{V}_u$ containing $N$ items, the objective of DeGRe is to generate an optimal sequence $l^*$ of length $L$ that maximizes the overall list value. To balance exploration depth and inference efficiency within the vast permutation space, DeGRe adopts an offline--online decoupled design. The framework comprises two core components: the Lookahead Evaluator (denoted as DeGRe-E) and the Online Generator (denoted as DeGRe-G). These components operate collaboratively across three stages: in the offline stage, the evaluator leverages complete data to mine high-quality lookahead sequences for constructing dense supervision signals; in the training stage, the generator internalizes lookahead planning capabilities through hybrid distillation; and in the online stage, the generator approximates the global optimum through a single efficient greedy decoding pass. Specifically, we detail the Lookahead Evaluator and dense supervision construction in Sections 4.1 and 4.2, introduce the Online Generator in Section 4.3, and elaborate on the optimization objective and inference strategy in Section 4.4.

\subsection{The Lookahead Evaluator}

We design the Lookahead Evaluator $E_\phi$ as a cumulative regression model based on a Causal Transformer, which estimates the cumulative value distribution of any subsequence $l_{1:t}$.

\textbf{Causal Sequence Encoder.} The input consists of $N_f$ feature fields mapped through embedding layers, including user embedding $\mathbf{e}_{user}$, item embedding $\mathbf{e}_{item}$, context embedding $\mathbf{e}_{ctx}$, and ranking score embedding $\mathbf{e}_{score}$. We replicate static features such as user and context along the sequence dimension to align with the item sequence, then stack all features to form the input tensor $\mathbf{E} \in \mathbb{R}^{L \times N_f \times d}$, where $d$ denotes the embedding dimension. Given the distributional differences across heterogeneous domains (e.g., user-side versus item-side)~\cite{Han2025MTGR}, we partition $\mathbf{E}$ into several heterogeneous feature groups and apply Group Layer Normalization (GLN) to normalize each group independently. Subsequently, we introduce inter-domain attention to generate enhanced representations $\tilde{\mathbf{E}}$. Finally, we flatten both the original features and enhanced representations, concatenate them with positional encoding $\mathbf{P} \in \mathbb{R}^{L \times d_{pos}}$, and project through a linear layer to obtain the Transformer input matrix $\mathbf{X} \in \mathbb{R}^{L \times d_{model}}$:
\begin{equation}
\mathbf{X} = \text{Linear}(\text{Flatten}(\mathbf{E}) \parallel \text{Flatten}(\tilde{\mathbf{E}}) \parallel \mathbf{P})
\end{equation}
where $\parallel$ denotes concatenation. This input matrix $\mathbf{X}$ is then fed into a Causal Transformer encoder with $N_E$ layers to capture contextual dependencies within the sequence, outputting the hidden state sequence $\mathbf{H} \in \mathbb{R}^{L \times d_{model}}$:
\begin{equation}
\mathbf{H} = [\mathbf{h}_1, \mathbf{h}_2, \dots, \mathbf{h}_L] = \text{TransformerEnc}(\mathbf{X})
\end{equation}

\textbf{Cumulative Regression Head.} We adopt cumulative regression~\cite{McCullagh1980Ordinal} to model the distribution of discrete metrics such as click counts. The model projects $\mathbf{H}$ through an MLP to obtain the value distribution matrix $\mathbf{O} \in \mathbb{R}^{L \times L}$, and applies the sigmoid function to compute the probability that the cumulative value $V$ reaches threshold $k$ ($1 \le k \le t$):
\begin{equation}
P(V \ge k | l_{1:t}) = \sigma(O_{t,k})
\end{equation}
where $O_{t,k}$ represents the unnormalized logit for the cumulative value reaching threshold $k$ at step $t$.

\textbf{Training Objective.} Training employs an ordered binary cross-entropy (BCE) loss~\cite{Frank2001Ordinal}. For a training sequence $l$ and its ground-truth cumulative value $y_t$ at each step, we construct a dense label matrix $\mathbf{Y} \in \{0,1\}^{L \times L}$, where $y_{t,k} = \mathbb{I}(y_t \geq k)$. The loss function is defined as:
\begin{equation}
\mathcal{L}_{Eval} = - \sum_{(l, y) \in \mathcal{D}} \frac{1}{L} \sum_{t=1}^L \sum_{k=1}^t \mathcal{L}_{BCE}(y_{t,k}, \sigma(O_{t,k}))
\end{equation}
Here, the upper limit of the inner summation $k \le t$ reflects the intrinsic constraint that the cumulative value cannot exceed the current step count, implemented through a lower triangular mask. Finally, we compute the expected cumulative value $\hat{V}(l_{1:t}) = \sum_{k=1}^t P(V \ge k | l_{1:t})$ based on the predicted distribution, which is used to guide generator training.

\begin{figure*}[t]
  \centering
  \includegraphics[width=\linewidth,height=0.34\textheight,keepaspectratio]{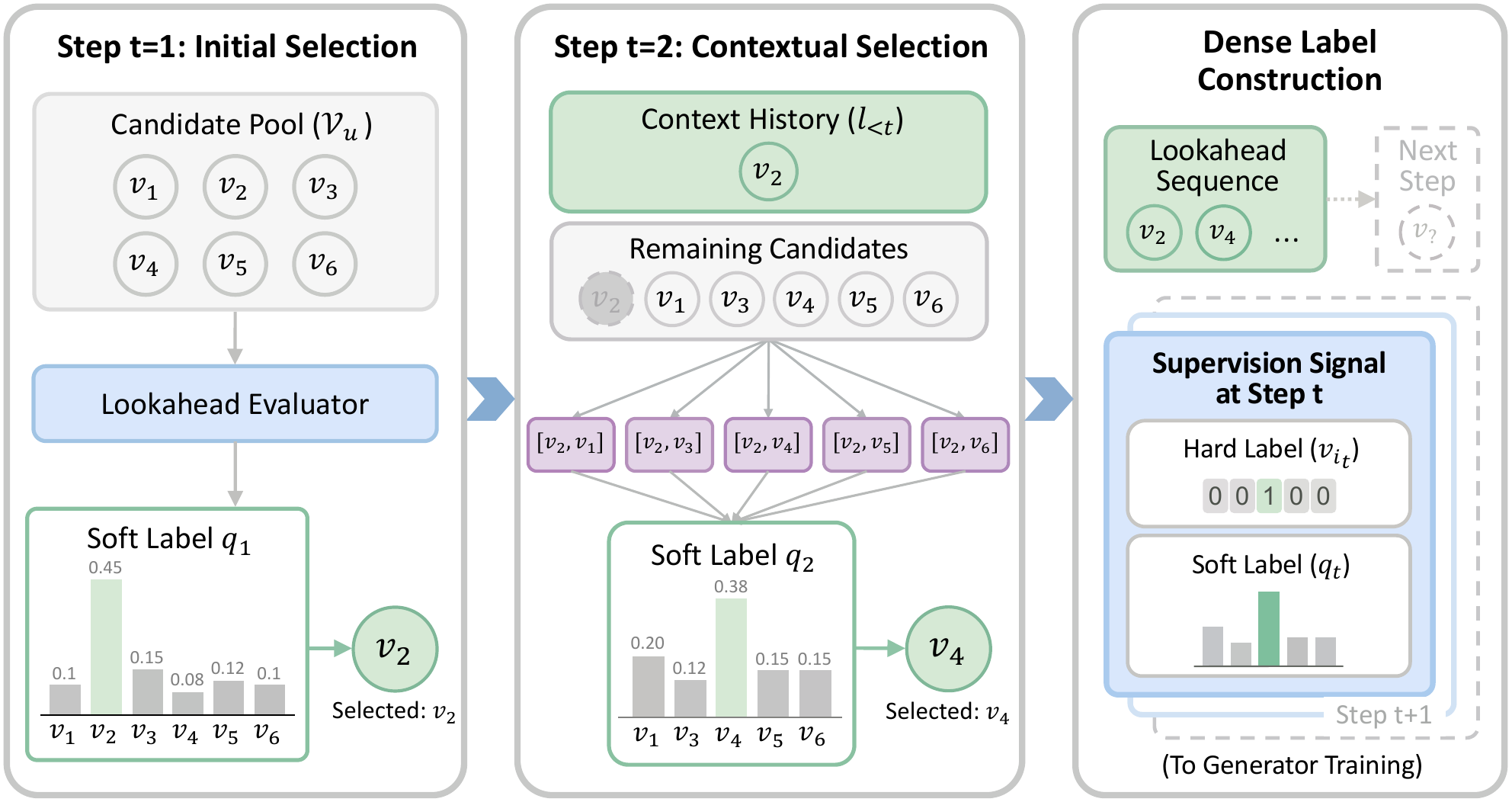}
  \caption{Illustration of the dense supervision construction process. At each step (e.g., $t=1, 2$), the Lookahead Evaluator determines the target decision (hard label $v_{i_t}$, such as $v_2, v_4$) and constructs soft labels ($q_t$) to preserve fine-grained ranking information. These dense signals, which combine deterministic and probabilistic information, are used to guide the training of the generator.}
  \label{fig:fig3}
\end{figure*}

\subsection{Dense Supervision Construction}

To overcome the limitations of heuristic labels in exploring the unexposed space, we leverage the trained evaluator $E_\phi$ to construct a dense supervision dataset in the offline stage.

\textbf{Lookahead Sequence Mining.} To actively mine high-value sequences in the unexposed space, we employ the Lookahead Evaluator $E_\phi$ to guide a beam search process with beam size $B$. As illustrated in Figure~\ref{fig:fig3} (using a single candidate path as an example), at each search step $t$, we expand the currently maintained subsequences by iterating over all remaining candidate items $\mathcal{V}_u \setminus l_{<t}$, and select the Top-$B$ paths with the highest cumulative value based on the value estimation $\hat{V}([l_{<t}; v])$ from the evaluator. This process continues until the sequence length reaches $L$, and the resulting lookahead sequence set $\mathcal{T}_{syn} = \{l^{(b)}\}_{b=1}^B$ constitutes potential optimal solutions from the perspective of the evaluator.

\textbf{Dense Label Construction.} To distill the planning capability of the Lookahead Evaluator into the generator, we construct hybrid supervision signals that combine deterministic and probabilistic information at each time step $t$. Specifically, for any lookahead sequence $l = [v_{i_1}, \dots, v_{i_L}]$ in the set $\mathcal{T}_{syn}$, we define the item $v_{i_t}$ selected at step $t$ as the hard label, serving as the target decision for the generator to fit. Meanwhile, to incorporate fine-grained ranking information from suboptimal alternatives, we construct a soft label distribution $q_t$ using the value estimations of the evaluator for the remaining candidate items:
\begin{equation}
q_t(v_i) = \frac{\exp(\hat{V}([l_{<t}; v_i]))}{\sum_{v_j \in \mathcal{V}_u \setminus l_{<t}} \exp(\hat{V}([l_{<t}; v_j]))}
\end{equation}
This distribution serves as a local regularizer, enabling the generator to perceive the relative merits of non-target items within the candidate space.

\textbf{Sequence Weighting.} Although $\mathcal{T}_{syn}$ contains potential optimal solutions, sequences within the set still exhibit value disparity. To reshape the supervision intensity according to lookahead value, we introduce a sequence weighting mechanism for the set $\mathcal{T}_{syn}$. For any lookahead sequence $l \in \mathcal{T}_{syn}$, the normalized weight is computed as:
\begin{equation}
w_l = \frac{\exp(\hat{V}(l) / \tau_w)}{\sum_{l' \in \mathcal{T}_{syn}} \exp(\hat{V}(l') / \tau_w)}
\end{equation}
During optimization, we multiply $w_l$ by $B$ to eliminate the influence of beam size on the loss magnitude. Combined with the temperature coefficient $\tau_w$ that controls the sharpness of the distribution, this mechanism ensures that gradient updates are predominantly driven by high-confidence sequences.

\textbf{Scalability of Offline Mining.} To ensure scalability for larger candidate sets, the search space can be effectively constrained by applying Top-$K$ pre-truncation at each step. Furthermore, since the $\mathcal{O}(N \times B)$ expanded sequences evaluated per step are strictly independent, they can be merged into a single GPU batch for parallel scoring. This design allows the exploration overhead to be flexibly controlled by adjusting the beam size $B$.

\subsection{Efficient Online Generator}

The Online Generator $G_\theta$ adopts a lightweight Encoder--Decoder architecture, designed to satisfy the strict constraints of high throughput and low latency in industrial recommendation systems.

\textbf{Bidirectional Candidate Encoder.} During the encoding phase, the generator has global visibility over the candidate set $\mathcal{V}_u$. To maximize the efficiency of knowledge transfer from the evaluator to the generator, the generator employs the same underlying feature processing architecture as the Lookahead Evaluator, including GLN and inter-domain attention. Based on the aligned features, we use an $N_G$-layer bidirectional Transformer to encode the candidate set in parallel, obtaining a context-aware candidate representation matrix $\mathbf{M} \in \mathbb{R}^{N \times d_{model}}$. This allows the model to explicitly capture competitive and complementary relationships among candidate items.

\textbf{User-Guided Causal Decoder.} Existing autoregressive models~\cite{Bello2018Seq2Slate, Lin2025GReF} typically initiate generation with a static \texttt{<BOS>} token, resulting in a lack of personalization for the first item decision. To address this, we map the user representation $\mathbf{e}_{user}$ through a linear projection to obtain the initial input embedding for the decoder:
\begin{equation}
\mathbf{e}_{start} = \mathbf{W}_p \mathbf{e}_{user} + \mathbf{b}_p
\end{equation}
Subsequently, we concatenate $\mathbf{e}_{start}$ with the target sequence along the temporal dimension, then concatenate positional encoding along the feature dimension and fuse them through a linear layer. The decoder body utilizes an $N_G$-layer Causal Transformer to model sequential dependencies, outputting a hidden state $\mathbf{h}_t^{dec}$ at each step $t$ that captures sequential information.

\textbf{Candidate-Constrained Decoding.} Unlike general language models that generate from a fixed vocabulary, we adopt a candidate-constrained decoding mechanism inspired by pointer networks~\cite{Vinyals2015PointerNet, Bello2018Seq2Slate}. This mechanism strictly confines the output space to the candidate set $\mathcal{V}_u$, which varies dynamically with the input. Specifically, the model uses the current decoding hidden state $\mathbf{h}_t^{dec}$ as the query vector and computes dot-product interactions with the entire candidate representation matrix $\mathbf{M}$ (serving as keys) to obtain the normalized selection probability:
\begin{equation}
P_\theta(v_i | l_{<t}) = \frac{\exp(\mathbf{h}_t^{dec} \mathbf{m}_i^\top)}{\sum_{v_j \in \mathcal{V}_u \setminus l_{<t}} \exp(\mathbf{h}_t^{dec} \mathbf{m}_j^\top)}
\end{equation}
where $\mathbf{m}_i \in \mathbb{R}^{d_{model}}$ is the row vector corresponding to candidate item $v_i$ in the encoder output matrix $\mathbf{M}$. This mechanism ensures that generated items are strictly confined to the candidate set $\mathcal{V}_u$ without repetition.

\subsection{Training Objective and Inference}

\textbf{Hybrid Distillation Objective.} The training objective of the generator is to fit the offline-constructed $\mathcal{T}_{syn}$ via a hybrid distillation loss. The total loss function is defined as:
\begin{equation}
\mathcal{L}_{Gen} = \sum_{l \in \mathcal{T}_{syn}} w_l \cdot \sum_{t=1}^L \left[ \underbrace{\mathcal{L}_{CE}(v_{i_t}, P_{\theta, t})}_{\text{Lookahead Imitation}} + \underbrace{\alpha \mathcal{L}_{KL}(q_t \| P_{\theta, t})}_{\text{Value Alignment}} \right]
\end{equation}
where $P_{\theta,t}$ denotes the selection probability distribution of the generator at step $t$, and the loss contribution of each sequence is weighted by its importance weight $w_l$. $\mathcal{L}_{CE}$ supervises the generator to fit the target decisions of the evaluator using the hard label $v_{i_t}$, achieving deterministic imitation of lookahead sequences; $\mathcal{L}_{KL}$ guides the generator to align with the complete value distribution over the candidate space through the soft label $q_t$, preserving fine-grained ranking information.

\textbf{Efficient Inference.} A core advantage of DeGRe lies in the efficiency of online inference. During online serving, the computationally expensive Lookahead Evaluator $E_\phi$ does not need to be deployed. The generator $G_\theta$ only needs to perform a single greedy decoding pass to produce the recommendation list. Given that the generated sequence length $L$ is typically short (e.g., $L < 10$), this process avoids the additional computational overhead introduced by the evaluator in traditional two--stage paradigms, thereby effectively satisfying the low-latency requirements of industrial real-time systems.

\section{Experiments}

\subsection{Experimental Setup}

\subsubsection{Datasets}

To validate the effectiveness of DeGRe, we conduct extensive experiments on both public benchmark datasets and large-scale industrial datasets. The statistics are summarized in Table~\ref{tab:datasets}.

\begin{itemize}

\item \textbf{ML-1M}\footnote{\url{https://grouplens.org/datasets/movielens/1m/}}. This is a widely used public benchmark for movie recommendation, derived from the one million version of the MovieLens dataset. It contains approximately 1 million rating records from about 6,000 users on roughly 3,700 movies. We transform the ratings into implicit feedback signals, where a rating $\geq 5$ is treated as a positive sample.

\item \textbf{Taobao Ad}\footnote{\url{https://tianchi.aliyun.com/dataset/56}}. This is a public dataset derived from the display advertising system of Taobao. Each sample contains user ID, timestamp, behavior type, and item attribute features. Following~\cite{Feng2021GRN, Feng2021PRS}, we construct reranking candidate sets by segmenting interactions into lists and mixing in negative samples to simulate the list input format of the reranking task.

\item \textbf{Taobao Flash Shopping.} This is an industrial-scale dataset from Taobao Flash Shopping, which serves over 100 million daily active users. Each sample corresponds to a user request, containing 12 candidate items and 6 exposed items, with click and conversion labels annotated. Additionally, each sample includes user, item, and context features, as well as scoring information from the ranking stage.

\end{itemize}

\begin{table}[h]
\setlength{\heavyrulewidth}{0.05em}
\caption{Statistics of datasets}
\label{tab:datasets}
\begin{tabular}{lccc}
\toprule
\textbf{Dataset} & \textbf{\# Users} & \textbf{\# Items} & \textbf{\# Records} \\
\midrule
\textbf{ML-1M} & 6,040 & 3,706 & 1,000,209 \\
\textbf{Taobao Ad} & 1,141,729 & 99,815 & 26,557,961 \\
\textbf{Taobao Flash Shopping} & 3,511,657 & 1,753,654 & 167,911,273 \\
\bottomrule
\end{tabular}
\end{table}

\begin{table*}[t]
\caption{Performance comparison of generators on the ML-1M, Taobao Ad, and Taobao Flash Shopping datasets. To ensure a fair comparison, all methods are evaluated by an independent external evaluator under a unified configuration.}
\label{tab:generator_perf}
\begin{tabular}{lccccccccc}
\toprule
\multirow{2}{*}{\textbf{Model}} & \multicolumn{3}{c}{\textbf{ML-1M}} & \multicolumn{3}{c}{\textbf{Taobao Ad}} & \multicolumn{3}{c}{\textbf{Taobao Flash Shopping}} \\
\cmidrule(lr){2-4} \cmidrule(lr){5-7} \cmidrule(lr){8-10}
 & \textbf{HR@1\%} & \textbf{HR@3\%} & \textbf{HR@10\%} & \textbf{HR@1\%} & \textbf{HR@3\%} & \textbf{HR@10\%} & \textbf{HR@1\%} & \textbf{HR@3\%} & \textbf{HR@10\%} \\
\midrule
\textbf{NAR4Rec} & 0.1866 & 0.2863 & 0.4271 & 0.0966 & 0.1810 & 0.3378 & 0.1342 & 0.2234 & 0.4150 \\
\textbf{GReF} & 0.1926 & 0.2916 & 0.4368 & 0.1013 & 0.1932 & 0.3647 & 0.1618 & 0.2554 & 0.4564 \\
\textbf{NLGR-G} & 0.1656 & 0.3067 & 0.5224 & 0.0916 & 0.1955 & 0.4028 & 0.1225 & 0.2386 & 0.4645 \\
\textbf{GoalRank} & 0.5920 & 0.7871 & 0.9386 & 0.4211 & 0.6400 & 0.8501 & 0.3553 & 0.5669 & 0.8164 \\
\textbf{DeGRe-G (B=1)} & 0.7486 & 0.9204 & 0.9874 & 0.5222 & 0.7650 & 0.9458 & 0.6947 & 0.8554 & 0.9639 \\
\textbf{DeGRe-G (B=8)} & \textbf{0.8910} & \textbf{0.9669} & \textbf{0.9934} & \textbf{0.7106} & \textbf{0.8881} & \textbf{0.9833} & \textbf{0.8872} & \textbf{0.9670} & \textbf{0.9962} \\
\bottomrule
\end{tabular}
\end{table*}

\subsubsection{Baselines}

Given the decoupled generation-evaluation architecture of DeGRe, we divide the most representative state-of-the-art methods into two groups for comprehensive performance analysis, comparing them against the generator and evaluator of DeGRe respectively.

\textbf{Generator Baselines.} This group aims to validate the capability of the generator to search for high-quality lists within the vast permutation space.

\begin{itemize}

\item \textbf{NAR4Rec}~\cite{Ren2024NAR4Rec}: A non-autoregressive generative model that combines non-likelihood training with a contrastive decoding mechanism.

\item \textbf{GReF}~\cite{Lin2025GReF}: A unified generative reranking framework that integrates Rerank-DPO \cite{Rafailov2023DPO}, a reinforcement learning (RL) alignment technique, with ordered multi-token prediction (OMTP).

\item \textbf{NLGR-G}~\cite{Wang2025NLGR}: The generator component of NLGR, which captures relative scoring gradients using neighbor lists as counterfactual signals and employs a sampling-based non-autoregressive mechanism for combinatorial optimization.

\item \textbf{GoalRank}~\cite{Zhang2025GoalRank}: A generative reranking framework based on the Group-Relative Optimization principle, which utilizes a reward model to construct the reference policy.

\end{itemize}

\textbf{Evaluator Baselines.} This group aims to validate the accuracy of the evaluator in sequence value estimation.

\begin{itemize}

\item \textbf{DeepFM}~\cite{Guo2017DeepFM}: A classic point--wise model that combines FM and DNN to capture both low-order and high-order feature interactions.

\item \textbf{PIER}~\cite{Shi2023PIER}: A context-aware prediction model that leverages an omnidirectional attention mechanism to capture fine-grained feature interactions and list item dependencies.

\item \textbf{NLGR-E}~\cite{Wang2025NLGR}: The evaluator component of NLGR, which introduces a D-Attention unit to decouple feature context for estimating overall list value.

\end{itemize}

\subsubsection{Metrics}

For offline experiments, following~\cite{Wang2025NLGR}, we design separate evaluation metric systems for the generator and evaluator given their different task objectives.

\textbf{Generator Metrics.} We adopt Hit Rate @ Top-K\% (HR@K\%) based on Monte Carlo sampling to evaluate the optimization capability of the generator within the permutation space $\Pi(\mathcal{V}_u)$. This metric measures whether the generated list $l_{gen}$ ranks within the top $K\%$ of a random comparison set $\mathcal{S}_u$ (where $|\mathcal{S}_u|=10,000$):
\begin{equation}
\text{HR}@K\% = \frac{1}{|\mathcal{D}|} \sum_{u \in \mathcal{D}} \mathbb{I} \left( \frac{\text{Rank}(l_{gen}, \mathcal{S}_u)}{|\mathcal{S}_u|} < \frac{K}{100} \right)
\end{equation}
where $\text{Rank}(\cdot)$ denotes the number of samples in $\mathcal{S}_u$ with value higher than $l_{gen}$. We report results for $K \in \{1, 3, 10\}$.

\textbf{Evaluator Metrics.} We adopt the following metrics to evaluate the value estimation accuracy of the evaluator:

\begin{itemize}

\item \textbf{R-AUC (Regression AUC):} We extend binary classification AUC to the continuous value domain to measure ranking capability. It is defined as the probability that the predicted relative order is consistent with the ground-truth order:
\begin{equation}
\text{R-AUC} = \frac{\sum_{i,j: y_i > y_j} \mathbb{I}(\hat{y}_i > \hat{y}_j)}{\sum_{i,j} \mathbb{I}(y_i > y_j)}
\end{equation}

\item \textbf{PCOC \& RMSE:} We use PCOC (the ratio of predicted mean to ground-truth mean) to evaluate model calibration, and RMSE to measure absolute prediction error.

\end{itemize}

It is worth noting that R-AUC and HR measure model performance from complementary dimensions. R-AUC reflects the discriminative capability of the evaluator, while HR validates the optimization capability of the generator. Together, they establish both the direction and depth of optimization, and any weakness in either component limits the final recommendation effectiveness. For online experiments, we adopt CTR, ORDER, and GMV as business evaluation metrics.

\subsubsection{Implementation Details}

We implement all models using PyTorch on NVIDIA A100 (80GB) GPUs. For model architecture, we configure the number of Transformer layers for the evaluator and generator as $N_E=6$ and $N_G=4$ respectively, with an embedding size of 16. Training uses the Adam optimizer with a learning rate of $5 \times 10^{-4}$ and a batch size of 1,024. The weight parameter for soft label distillation is set to $\alpha = 0.01$. For each dataset, we adopt a reranking setting of selecting 6 items from 12 candidates, resulting in a full permutation search space of approximately $P(12, 6) \approx 6.6 \times 10^5$. To strictly follow temporal causality, we uniformly adopt the leave-one-out strategy for data splitting, using the last session of each user as the test set and the remaining sessions for training. All experiments are repeated five times with different random seeds, and we report the average results.

\subsection{Offline Experiments}

\subsubsection{Generator Performance}

Table~\ref{tab:generator_perf} presents the offline experimental results of the generator. We first observe that GoalRank, which is based on reward mechanisms, outperforms NAR4Rec across all metrics. NAR4Rec relies solely on MLE to fit exposed data, and this validates the value of exploration in the unexposed space. Building on this, DeGRe outperforms baseline models across all metrics on all three datasets. Notably, even under the weak supervision setting ($B=1$), DeGRe already surpasses the strongest baseline, GoalRank. When offline exploration is enhanced ($B=8$), DeGRe achieves HR@1\% of 89.10\%, 71.06\%, and 88.72\% on ML-1M, Taobao Ad, and Taobao Flash Shopping, respectively. This performance represents absolute improvements of 29.90\%, 28.95\%, and 53.19\% over the state-of-the-art methods, respectively. This consistent performance across datasets validates the robustness of DeGRe and its generalization capability in non-e-commerce domains.

\subsubsection{Evaluator Performance}

Table~\ref{tab:evaluator_perf} presents the performance of the evaluator on the Taobao Flash Shopping dataset. We observe that list--wise models such as PIER significantly outperform the point--wise model DeepFM in terms of R-AUC, validating the importance of capturing contextual dependencies for accurately discriminating sequence quality. Building on this, DeGRe-E outperforms all baselines across all metrics, with PCOC reaching 0.9932. This result validates the effectiveness of the cumulative regression mechanism in fitting discrete distributions and calibrating predictions.

\begin{table}[h]
\caption{Performance comparison of evaluators on the Taobao Flash Shopping dataset.}
\label{tab:evaluator_perf}
\centering
\setlength{\tabcolsep}{4pt}
\begin{tabular}{@{}lcccc@{}}
\toprule
\textbf{Metric} & \textbf{Point--wise} & \multicolumn{2}{c}{\textbf{List--wise}} & \textbf{Cumulative} \\
 & \textbf{DeepFM} & \textbf{PIER} & \textbf{NLGR-E} & \textbf{DeGRe-E} \\
\midrule
R-AUC $\uparrow$ & 0.6979 & 0.7041 & 0.7036 & \textbf{0.7090} \\
PCOC $\rightarrow 1$ & 1.0392 & 0.9132 & 0.8828 & \textbf{0.9932} \\
RMSE $\downarrow$ & 0.4985 & 0.4966 & 0.4973 & \textbf{0.4946} \\
\bottomrule
\end{tabular}
\end{table}

\subsection{Ablation Study}

To validate the effectiveness of different components in DeGRe, we construct four ablation variants on the Taobao Flash Shopping dataset under the $B=2$ setting.

\begin{itemize}

\item \textbf{w/o Soft Label}: This variant removes the soft label distillation loss and retains only hard labels for training to verify the role of fine-grained ranking information.

\item \textbf{w/o Seq Weighting}: This variant removes the weight computation for lookahead sequences to assess the impact of sequence value weighting, meaning all sequences are treated equally.

\item \textbf{w/o Hard Label}: This variant removes the hard label loss to verify the necessity of lookahead sequence imitation, and no longer forces the generator to reproduce the optimal sequences from the evaluator.

\item \textbf{Exposure Only}: Serving as a baseline lower bound, this variant completely removes the lookahead planning module and trains the generator directly on historical exposure data.

\end{itemize}

\begin{table}[h]
\caption{Ablation results on the Taobao Flash Shopping dataset.}
\label{tab:ablation}
\begin{tabular}{lcc}
\toprule
\textbf{Variant} & \textbf{HR@1\%} & \textbf{HR@10\%} \\
\midrule
\textbf{DeGRe (B=2)} & \textbf{0.7910} & \textbf{0.9852} \\
w/o Soft Label & 0.7896 & 0.9856 \\
w/o Seq Weighting & 0.7856 & 0.9846 \\
w/o Hard Label & 0.2107 & 0.5631 \\
Exposure Only & 0.1577 & 0.4353 \\
\bottomrule
\end{tabular}
\end{table}

The results are shown in Table~\ref{tab:ablation}. From the experimental results, we have the following findings: i) The variant without hard label exhibits the most significant drop in HR@1\% (58.03\%) and performs only marginally better than exposure only, indicating that high-quality lookahead sequences mined via beam search form the foundation of DeGRe performance. ii) The variant without sequence weighting shows a decrease of 0.54\% in HR, indicating that value-aware weighting helps the model focus on high-confidence, high-quality sequences. iii) The variant without soft label also shows a decrease in HR (0.14\%), indicating that the distributional information provided by soft labels assists the model in better approximating the optimal solution.

\subsection{Hyperparameter Analysis}

\begin{figure}[t]
  \centering
  \includegraphics[width=\linewidth]{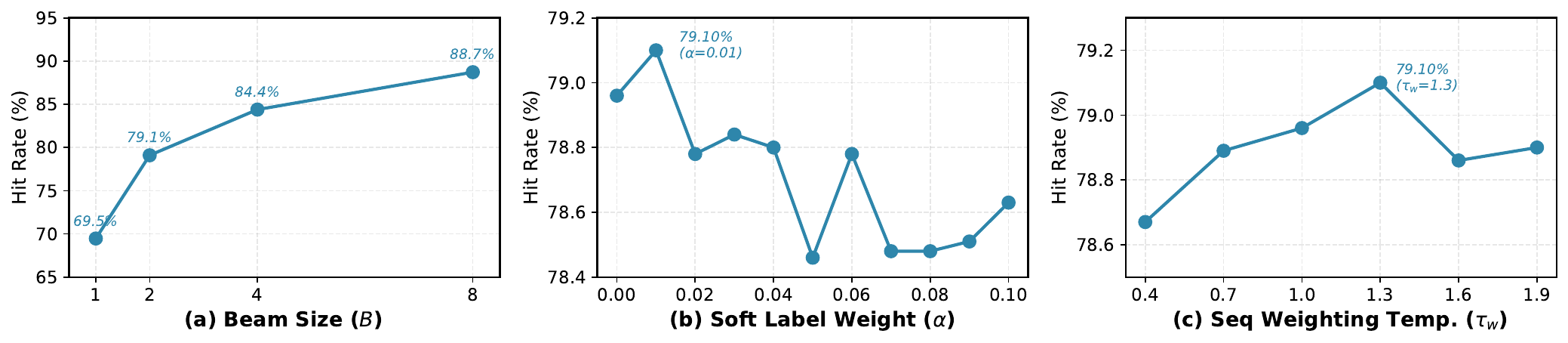}
  \caption{Hyperparameter sensitivity analysis on the Taobao Flash Shopping dataset (metric: HR@1\%). (a) Effect of the offline search beam size $B$; (b) Effect of the soft label weight $\alpha$; (c) Effect of the sequence weighting temperature $\tau_w$.}
  \label{fig:fig4}
\end{figure}

We analyze the sensitivity of three key hyperparameters in DeGRe: the beam size $B$ for offline search, the soft label weight $\alpha$, and the sequence weighting temperature $\tau_w$. Figure~\ref{fig:fig4} illustrates the trends in HR@1\%, from which we have the following findings: i) As $B$ increases, the Online Generator's performance improves significantly but with diminishing marginal returns (Figure~\ref{fig:fig4}a). This validates that a wider offline search space mines higher-quality lookahead sequences. ii) Performance first increases then decreases with $\alpha$, peaking at $\alpha=0.01$ (Figure~\ref{fig:fig4}b). This indicates that soft labels serve as an effective auxiliary regularizer by introducing fine-grained ranking information across the candidate space. However, an excessively large $\alpha$ leads to over-smoothing of the distribution, interfering with the main task's optimization. iii) Performance exhibits an inverted U-shaped pattern with $\tau_w$ (Figure~\ref{fig:fig4}c), indicating that a moderate temperature balances distinguishing high-value sequences with maintaining training stability.

\subsection{Online A/B Test}

We deploy DeGRe in the homepage recommendation scenario of Taobao Flash Shopping and conduct an 8-day online experiment with 2\% of live traffic under the A/B testing framework. In the experimental setup, we introduce PRM~\cite{Pei2019PRM} as an additional baseline alongside the base strategy (a multi-objective fusion point--wise model). As a classic single-stage reranking model in industry, PRM provides a key reference for evaluating the advantages of the generative paradigm.

\subsubsection{Overall Performance}

Table~\ref{tab:online_ab} presents the overall results of the online A/B test. Compared to the base strategy, DeGRe achieves substantial improvements of +2.85\%, +2.14\%, and +3.75\% in CTR, ORDER, and GMV, respectively. This indicates that the optimization of overall list utility through generative reranking effectively translates into actual transaction growth. Compared to PRM, DeGRe further improves ORDER and GMV by 1.0\% and 2.99\%, respectively, highlighting the advantage of the generative paradigm over single-stage greedy strategies in optimizing global list utility. Thanks to the offline--online decoupled design, the evaluator does not need to be deployed in the online stage, and the computational overhead of autoregressive generation is effectively controlled. The average inference latency of DeGRe increases by only 14.8 ms, meeting the low-latency requirements of large-scale real-time systems.

\begin{table}[h]
\caption{Online A/B test results on Taobao Flash Shopping}
\label{tab:online_ab}
\centering
\setlength{\tabcolsep}{5pt}
\begin{tabular}{lcccc}
\toprule
\textbf{Model} & \textbf{CTR} & \textbf{ORDER} & \textbf{GMV} & \textbf{Cost (ms)} \\
\midrule
Base & 0.00\% & 0.00\% & 0.00\% & - \\
PRM & +0.73\% & +1.14\% & +0.76\% & +6.2 \\
\textbf{DeGRe} & \textbf{+2.85\%} & \textbf{+2.14\%} & \textbf{+3.75\%} & +14.8 \\
\bottomrule
\end{tabular}
\end{table}

\subsubsection{Robustness Analysis}

\begin{figure}[H]
  \centering
  \includegraphics[width=\linewidth]{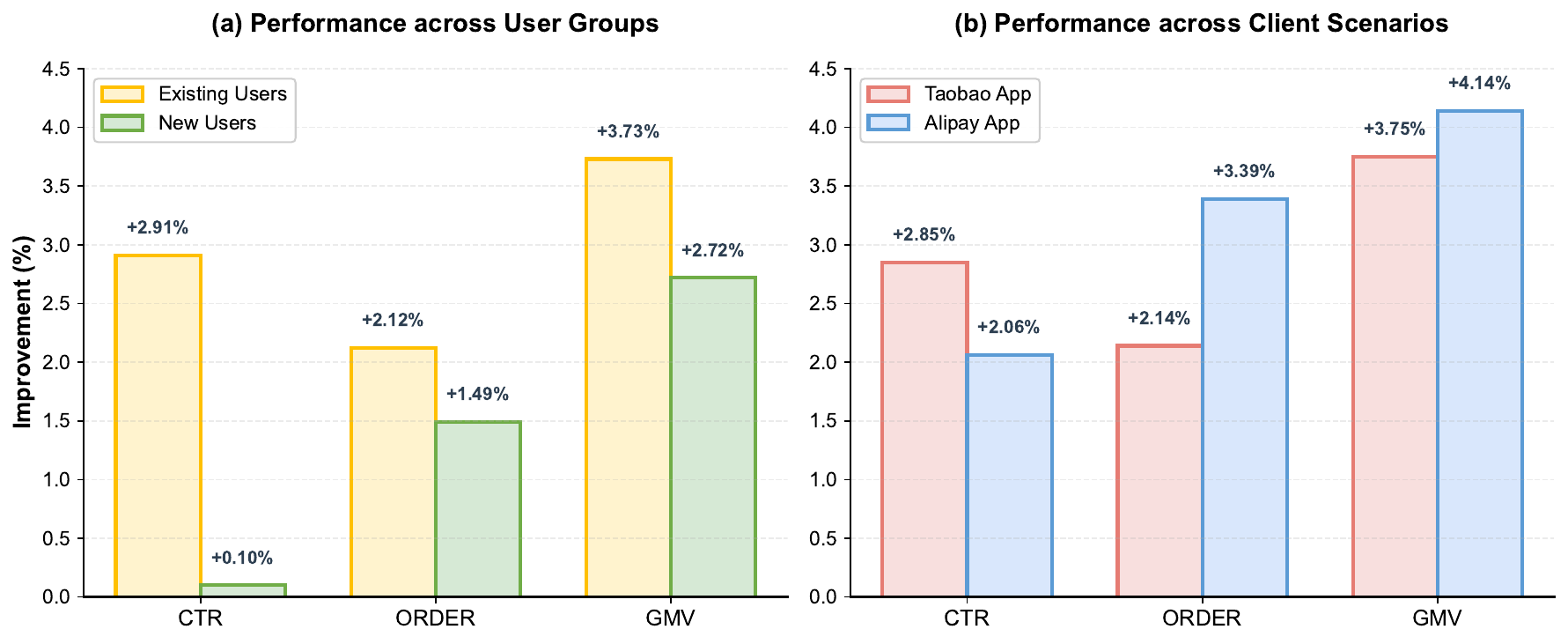}
  \caption{Robustness analysis of online A/B testing for DeGRe. (a) Performance comparison across different user groups; (b) Performance comparison across different client scenarios.}
  \label{fig:fig5}
\end{figure}

To validate the robustness of DeGRe, we further analyze the performance of the model across different user groups and client scenarios. The results are shown in Figure~\ref{fig:fig5}. For existing users with rich historical behaviors, the model achieves a GMV improvement of +3.73\%, while for new users with sparse behaviors, DeGRe still achieves a GMV increase of +2.72\%. This indicates that DeGRe generalizes well across user groups with different historical behavior densities. For different client scenarios, DeGRe demonstrates robust cross-scenario generalization capability. In addition to achieving significant gains on the Taobao App, the model also achieves a GMV improvement of +4.14\% and an order volume increase of +3.39\% on Alipay, which has a considerably different traffic distribution.

\section{Conclusion}

This paper proposes DeGRe, a novel framework for efficient generative reranking via dense supervision. DeGRe adopts an offline--online decoupled design: in the offline stage, a cumulative-regression-based Lookahead Evaluator actively mines high-value sequences through beam search; in the training stage, the evaluator's instantaneous value estimations are transformed into dense supervision signals and distilled into a lightweight Online Generator. This mechanism enables the generator to internalize lookahead planning capabilities, allowing it to approximate the global optimum with only a single greedy decoding step during online inference. Extensive offline experiments and online A/B tests demonstrate that DeGRe significantly outperforms existing reranking baselines, and it has been successfully deployed in Taobao Flash Shopping.

\section*{Acknowledgments}

This research was supported by The National Nature Science Foundation of China (Grant Nos: 62432014, 62402417, 62273301, 62273302), in part by ``Pioneer'' and ``Leading Goose'' R\&D Program of Zhejiang (Grant No. 2025C02026), in part by the Key R\&D Program of Ningbo (Grant Nos: 2024Z115, 2025Z035), in part by Yongjiang Talent Introduction Programme (Grant No: 2023A-197-G).

\bibliographystyle{ACM-Reference-Format}
\balance
\bibliography{references}

\end{document}